\documentclass[lnicst]{svmultln}
\usepackage{makeidx}  

\usepackage{caption}
\usepackage{graphicx}
\usepackage{url}
\usepackage[autostyle]{csquotes}
\usepackage{amssymb,amsmath}
\usepackage{epsfig}
\usepackage{epstopdf}
\usepackage{setspace}
\usepackage{amsthm}
\usepackage{enumerate}
\usepackage{tikz}
\usetikzlibrary{shapes,arrows}
\usepackage{algorithm}
\usepackage{algcompatible}
\usepackage{float}
\usepackage{algpseudocode}
\usepackage{pifont}
\begin{document}
\mainmatter              
\title{Nash Equilibrium in Social Media}
\titlerunning{Nash Equilibrium Seeking}  
%
\author{Farzad Salehisadaghiani}
\authorrunning{F. Salehisadaghiani}   
%
%
\institute{Department of Electrical and Computer Engineering, University of Toronto, 10 King's College Road, Toronto, ON, M5S 3G4, Canada
\email{farzad.salehisadaghiani@mail.utoronto.ca}}

\maketitle              
\vspace{1cm}\section{Introduction}
In this work, we investigate an application of a Nash equilibrium seeking algorithm in a social network. In a networked game each player (user) takes action in response to other players' actions in order to decrease (increase) his cost (profit) in the network.\\

We assume that the players' cost functions are not necessarily dependent on the actions of all players. This is due to better mimicking the standard social media rules. A communication graph is defined for the game through which players are able to share their information with only their neighbors. We assume that the communication neighbors  necessarily affect the players' cost functions while the reverse is not always true.\\

In this game, the players are only aware of their own cost functions and actions. Thus, each of them maintains an estimate of the others' actions and share it with the neighbors to update his action and estimates.\\

A Nash equilibrium is a point that no player can unilaterally deviate his action to gain a better profit while the others keep their actions unchanged. In this work, we do not study the various algorithms within this framework, but instead, we analyze a specific application of these algorithms in social networks. Interested readers may refer to \cite{salehisadaghiani2016distributed,salehisadaghiani2016nondoubly,salehisadaghiani2016distributedifac,ssalehisadaghiani2016distributed,salehisadaghiani2014nash,salehisadaghiani2017generalizedarxiv} for a detailed explanation of such algorithms.  
\newpage
\subsection{Social Media Behavior}
In this section we aim to investigate social networking media for users' behavior. In such media like Facebook, Twitter and Instagram users are allowed to follow (or be friend with) the other users and post statuses, photos and videos or also share links and events. Depending on the type of social media, the way of communication is defined. For instance, in Instagram, friendship is defined unidirectional in a sense that either side could be only a follower and/or being followed.\\

Recently, researchers at Microsoft have been studying the behavioral attitude of the users of Facebook as a giant and global network \cite{Forbes}. This study can be useful in many areas e.g. business (posting advertisements) and politics (posting for the purpose of presidential election campaign).\\

Generating new status usually comes with the cost for the users such that if there is no benefit in posting status, the users don't bother to generate new ones. In any social media drawing others' attention is one of the most important motivation/stimulation to post status \cite{goel2012game}. \textbf{Our objective is to find the optimal rate of posting status for each selfish user to draw more attention in his network}. In the following, we make an information/attention model of a generic social media \cite{goel2012game}, define a way of communication between users via a digraph $G_C$, and mark the interactions between users by an interference digraph $G_I$.\\

Consider a social media network of $N$ users. Each user $i$ produces $x_i$ unit of information such that the followers can see in their news feeds. The users' communication network is defined by a strongly connected digraph $G_C$ in which \raisebox{.5pt}{\textcircled{\raisebox{-.9pt} {$i$}}}$\rightarrow$\raisebox{0.5pt}{\textcircled{\raisebox{-.5pt} {$j$}}} means $j$ is a follower of $i$ or $j$ receives $x_i$ in his news feed. We also assume a strongly connected interference digraph $G_I$ to show the influence of the users on the others. We assume that each user $i$'s cost function is not only dependent on the users he follows, but also by the users that his followers follow.\\

The cost function of user $i$ is denoted by $J_i$ and consists of three parts:  
\begin{enumerate}
	\item $C_i(x_i)$: A cost that user $i$ pays to produce $x_i$ unit of information.
	\begin{equation*}\label{facebook_Produce_Cost}
	C_i(x_i):=h_ix_i,
	\end{equation*}
	where $h_i>0$ is a user-specific parameter.
	\item $f_i^1(x)$: A differentiable, increasing and concave utility function of user $i$ from receiving information from his news feed with $f_i^1(\textbf{0})=0$.
	\begin{equation*}\label{facebook_information_utility}
	f_i^1(x):=L_i\sqrt{\sum_{j\in N_C^\text{in}(i)}q_{ji}x_j},
	\end{equation*}
	where $q_{ji}$ represents follower $i$'s interest in user $j$'s information and $L_i>0$ is a user-specific parameter.\\
	\item $f_i^2(x)$: An incremental utility function that each user $i$ obtains from receiving attention in his network with $f_i^2(x)|_{x_i=0}=0$. Specifically, this function targets the amount of attention that each follower pays to the information of other users in his news feed.
	\begin{eqnarray}\label{facebook_attention_utility}
	\hspace{-1.4cm}&&f_i^2(x)=\sum_{l:i\in N_C^\text{in}(l)}\!L_l\Big(\sqrt{\!\sum_{j\in N_C^\text{in}(l)}q_{jl}x_j}-\sqrt{\!\sum_{j\in N_C^\text{in}(l)\backslash\{i\}}q_{jl}x_j}\Big).\nonumber
	\end{eqnarray}
	In fact, $f_i^2(x)$ computes the total profit of the followers of user $i$ from receiving information from the users excluding $i$ and subtract it from the one that includes $i$.
\end{enumerate}

The total cost function for user $i$ is then $J_i(x)=C_i(x_i)-f_i^1(x)-f_i^2(x)$.
For this example, we consider 5 users in the social media whose network of followers $G_C$ is given in Fig. 2. (a). From $G_C$ and taking $J_i$ into account, one can construct $G_I$ (Fig. 2. (b)) in a way that the interferences among users are specified.
\begin{figure}
	\vspace{-2cm}
	\hspace{-9.5cm}
	\centering
	\includegraphics [scale=1]{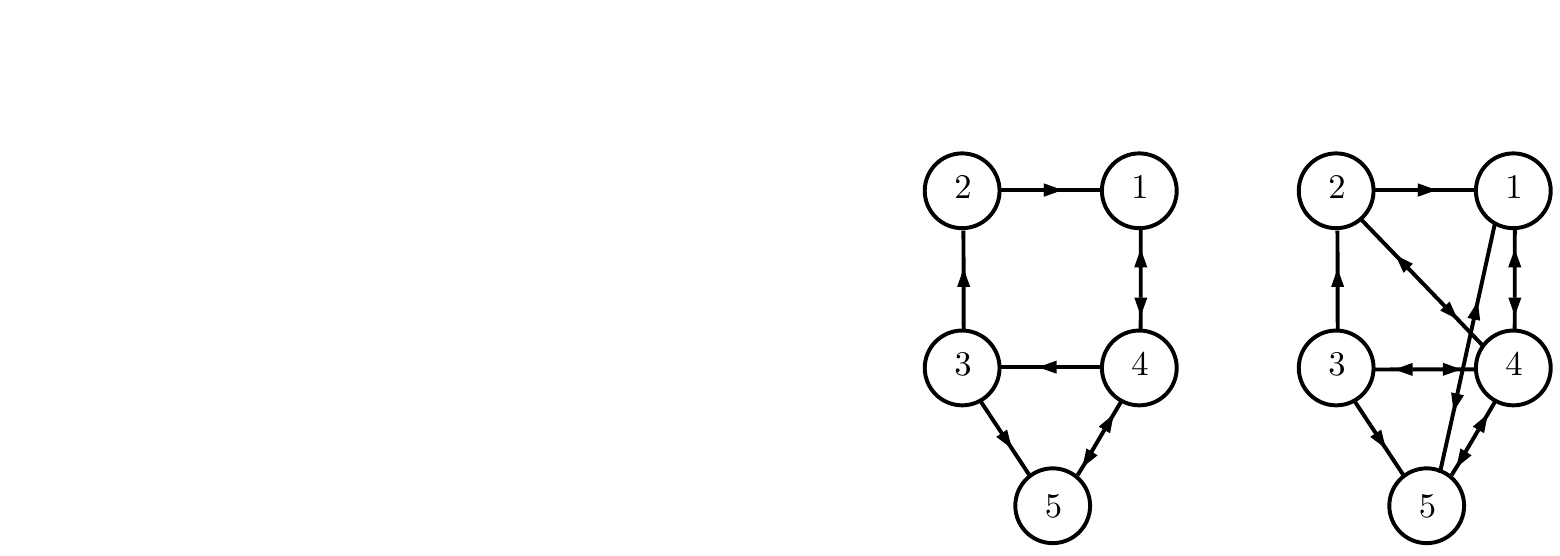}
	\label{fig11111}
	\caption{(a) $G_C$ (b) $G_I$.}
\end{figure}  
Note that this is a reverse process of the one discussed in \cite{ssalehisadaghiani2016distributed} because $G_C$ is given as the network of followers and $G_I$ is constructed from $G_C$. For the particular networks in Fig. 2, assumptions in \cite{ssalehisadaghiani2016distributed} hold. We then employ the algorithm in \cite{ssalehisadaghiani2016distributed} to find an NE of this game for $h_i=2$ and $L_i=1.5$ for $i\in V$, and $q_{41}=q_{45}=1.75$, $q_{32}=q_{43}=2$ and the rest of $q_{ij}=1$.
\begin{figure}
	\vspace{-4cm}
	\hspace{1cm}
	\includegraphics [scale=0.5]{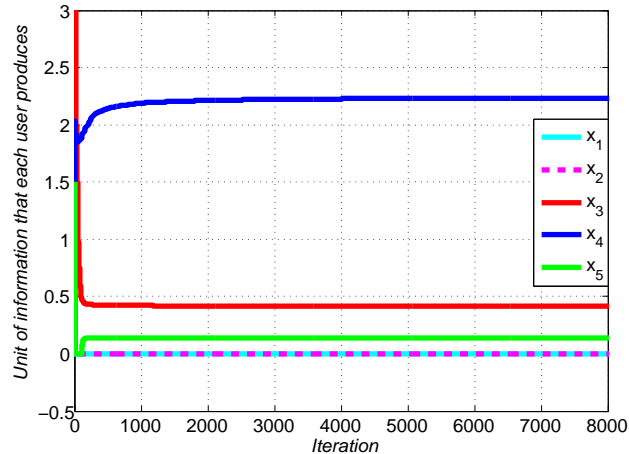}
	\label{fig11111}
	\vspace{-4cm}
	\caption{Convergence of the unit of information that each user produces to a NE over $G_C$.}
\end{figure} 
\subsection{Analysis}
In this section we analyze the NE $x^*=[0,0,0.42,2.24,0.14]^T$. One can realize that from $G_C$ in Fig. 2 (a), user 4 has 3 followers (users 1, 3 and 5), user 3 has 2 followers (users 2 and 5) and the rest has only 1 follower. It is straightforward to predict that users 4 and 3 could draw more attentions due to their more number of followers which is end up having less cost. This let them to produce more information ($x_4^*\geq x_3^*\geq x_{j\in\{1,2,5\}}^*$). On the other hand, user 5 receives $x_4$ and $x_3$ from his news feed which ends up having greater payoff than users 1 and 2 from perceiving information. This is why $x_5^*\geq x_{j\in\{1,2\}}^*$.
%
%
%
%
%
%
%
%
%
%
\bibliographystyle{splncs03}
\bibliography{ref}
\end{document}